\documentclass[hyper]{JHEP3}

\usepackage{amsmath}
\usepackage{amssymb}
\usepackage{graphicx}

\newcommand{\bZ}{\mathbb{Z}}
\newcommand{\cN}{\mathcal{N}}
\newcommand{\cM}{\mathcal{M}}
\newcommand{\cF}{\mathcal{F}}
\newcommand{\cZ}{\mathcal{Z}}
\newcommand{\Tr}{\mathrm{Tr\,}}
\newcommand{\ov}{\overline}

\title{D-brane instantons and matrix models}

\author{I\~naki Garc\'ia-Etxebarria\\
  Department of Physics and Astronomy, University of Pennsylvania,\\
  Philadelphia, PA 19104-6396, USA\\
  E-mail: \email{inaki@sas.upenn.edu}}

\abstract{We point out that in some situations it is possible to use
  matrix model techniques a la Dijkgraaf-Vafa to perturbatively
  compute D-brane instanton effects. This provides an explanation in
  terms of stringy instantons of the results in hep-th/0311181. We
  check this proposal in some simple scenarios. We point out some
  interesting consequences of this observation, such as the fact that
  it gives a perturbative way of computing stringy multi-instanton
  effects. It also provides a further interpretation of D-brane
  instantons as residual instantons of higgsed supergroups.}

\preprint{UPR-1200-T}

\begin{document}

\section{Introduction}

In a series of remarkable papers
\cite{Dijkgraaf:2002fc,Dijkgraaf:2002vw,Dijkgraaf:2002dh}, Dijkgraaf
and Vafa have argued that the exact low energy superpotential for a
class of gauge theories can be computed by taking the planar limit of
a matrix model. This matrix model is determined in a simple way by the
deconstruction \cite{Dijkgraaf:2003xk} of the original supersymmetric
theory. It has a set of (bosonic) matrix valued fields in one to one
correspondence with the superfields of the original theory, and a
potential given by the superpotential of the original theory. For
completeness, we include a short review of the Dijkgraaf-Vafa results
in Section~\ref{sec:DV-review}.

\medskip

Remarkably, even if the effective superpotential is obtained from the
planar limit of the matrix model, it is the low energy superpotential
for finite $N$ in the original theory (with $N$ not necessarily
large). A particularly interesting case is that in which $N$ is
small. This case led originally to some puzzles, since for some
particular cases the results obtained from the matrix model and the
pure gauge theory seemed to disagree. The situation was clarified in
\cite{Aganagic:2003xq,Intriligator:2003xs}, where it was argued that
the matrix model results might indeed differ from the field theory
expectations for gauge factors of low rank, but the disagreement could
be attributed to the choice of a UV definition of the gauge theory:
the matrix model gives results for the definition of the gauge theory
corresponding to the particular geometric embedding of the gauge
theory in string theory chosen by Dijkgraaf and Vafa.

\medskip

In \cite{Aganagic:2003xq,Intriligator:2003xs} the UV completion of the
gauge theory was defined by its embedding in a (higgsed) supergroup of
high enough rank. At the level of F-terms this definition of the gauge
theory coincides with the usual gauge theory one, but it has the
advantage that it is well defined in the UV. It is also very natural
from the stringy point of view, where it can be understood as adding
to the original DV embedding of the gauge theory an infinite set of
brane-antibrane pairs with trivial K-theory charge.

\medskip

The purpose of this note is to point out, in the simplest possible set
of examples, that it is also possible to understand this discrepancy
between gauge theory and matrix models as coming from stringy D-brane
instanton effects
\cite{Blumenhagen:2006xt,Ibanez:2006da,Florea:2006si}. Our claim is
that the matrix model computation includes the contribution of D-brane
instantons to the low energy superpotential of the string theory
realization of the gauge theory.

\medskip

More precisely, we claim that it is possible to obtain the
contributions of D-brane instantons to the effective superpotential by
studying the planar limit of the associated matrix model. In the
conclusions we include some preliminary remarks about the
interpretation of the non-planar contributions, and we also make some
comments on how could it be possible to generalize our results to some
geometries beyond the Dijkgraaf-Vafa setup.

\medskip

An interesting consequence of this result is that it gives a
relatively easy and systematic way of computing perturbatively stringy
multi-instanton effects, at least when a matrix model description of
the system is available. These multi-instanton effects are typically
rather involved to study using stringy instanton calculus, so the
perturbative approach based on the matrix model will be a welcome
tool.

\medskip

We can find some other interesting implications of the result. There
are by now different ways of computing exotic D-brane instanton
effects by modifying the system such that it admits a gauge theory
interpretation while still allowing to extract the relevant
information. Let us mention as examples
\cite{Aharony:2007pr,GarciaEtxebarria:2007zv,Amariti:2008xu}, in which
Seiberg duality (as in the duality cascade) is used to turn D-brane
instantons into conventional gauge non-perturbative effects. Another
example is \cite{Krefl:2008gs}, where Seiberg duality is combined with
higgsing.

\medskip

From this perspective, the matrix model viewpoint offers yet another
embedding of D-brane instanton effects into gauge theory. It is given
in terms of the supergroup construction of \cite{Aganagic:2003xq},
which essentially consists of adding infinite brane-antibrane pairs to
the system. In this setup stringy D-brane instanton effects are
identified with residual instanton effects of the higgsed supergroup.

\medskip

Some of the constructions above also admit a nice interpretation in
terms of matrix models. For example, as explained in
\cite{Dijkgraaf:2002dh}, Seiberg duality in the case of ADE quiver
theories can be understood as Weyl reflections of the quiver diagram,
and the matrix model results are essentially insensitive to this Weyl
reflection. Similarly, as described in \cite{Dijkgraaf:2003xk}, the
F-terms computed by the matrix model are insensitive to where we are
in the duality cascade, so we can do our calculations either ``up the
cascade'', where everything can be formulated in terms of gauge
theory, or at the bottom of the cascade, where we have exotic D-brane
instanton effects with no gauge theory interpretation. Let us mention
that \cite{Dijkgraaf:2003xk} also provides a nice intermediate
viewpoint between the brane-antibrane configuration and the matrix
model in the form of the {\em super}matrix model description, which
turns out to be rather useful for understanding the physics of these
duality cascades in the context of matrix models.

\medskip

We would also like to point out that very similar considerations to
some of the ones we discuss here have already appeared in
\cite{Aganagic:2007py}. In that paper, the geometric transition is
used to compute non-perturbative effects due to stringy D-brane
instantons. The matrix model and the geometric transition approaches
are very closely related, and they can be shown to give the same
result in the planar limit, which determines the superpotential. This
is, nevertheless, a large $N$ effect\footnote{We hope that the
  terminology we use does not lead to confusion. The $N$ here refers
  to the rank of the matrices in the matrix model, and has nothing to
  do with the ranks of the physical gauge group. We review this point
  in Section~\ref{sec:DV-review}.}, and we expect that we have to use
the matrix model for computing higher derivative interactions in the
effective superpotential. If we restrict ourselves to the
superpotential, and thus the planar limit, the matrix model provides a
different but equivalent viewpoint of the discussion in
\cite{Aganagic:2007py}.

\medskip

We feel that the previous reasons make it worth pointing out the
explicit connection between D-brane instantons and matrix models. It
is, in any case, a subtle and interesting connection between two rich
fields of study.

\medskip

Let us mention that in this note we will be checking the proposed
connection only in the simplest possible scenario: we will be
computing one-instanton superpotential contributions for quiver gauge
theories admitting a matrix model description a la Dijkgraaf and
Vafa. This leaves many issues to be studied, and we hope to come back
to them in future work.

\medskip

The remainder of this note is organized as follows. We start in
Section~\ref{sec:background} by summarizing the main points of the
Dijkgraaf-Vafa correspondence and D-brane instanton calculus. In
Section~\ref{sec:low-rank} we see how the D-brane instanton and matrix
model results agree for the simplest one-instanton sector in conifold
geometries. In Section~\ref{sec:quiver} we study a more interesting
and involved example, the abelian orbifold of the
conifold. Section~\ref{sec:conclusions} contains our conclusions and
some discussion of interesting open problems.

\section{Background material}
\label{sec:background}

\subsection{The Dijkgraaf-Vafa correspondence}
\label{sec:DV-review}

Let us start by reviewing the main results of Dijkgraaf and Vafa
\cite{Dijkgraaf:2002fc,Dijkgraaf:2002vw,Dijkgraaf:2002dh}. Since we
will be dealing with stringy instantons we concentrate on the stringy
derivation and interpretation of the results, although the same
results can also be derived in pure field theory
\cite{Cachazo:2002ry,Dijkgraaf:2002xd,Ferrari:2002jp}.

\subsection*{Gauge theory}

The configurations we will be concerned with are stringy embeddings of
$\cN=1$ gauge theories with a mass gap, the lightest states being
glueball superfields\footnote{The existence of the mass gap, and
  glueballs being the correct low energy degrees of freedom, are
  statements which are very hard to show rigorously. Here we just
  assume them to hold.}. The canonical example is a $\cN=1$ theory
obtained from $\cN=2$ $U(N)$ SYM with no flavors and a potential for
the adjoint superfield:
\begin{equation}
  \label{eq:W-adjoint}
  \int d^2\theta \, \Tr W(\Phi)
\end{equation}
This superpotential breaks down the supersymmetry to $\cN=1$. We take
$W(z)$ to be a polynomial function of degree $n+1$. We will denote the
dynamical scale of this theory by $\Lambda$.

The classical vacua of these theory are obtained by extremizing the
superpotential $W$. This superpotential has $n$ extrema $a_i$, let us
indicate this by writing the derivative of $W$ as $W'(z) = g
\prod_{i=1}^n(z-a_i)$. A classical solution of $\Tr W'(\Phi) = 0$ is
obtained by considering a partition of the $N$ eigenvalues of $\Phi$
into the $n$ extrema:
\begin{equation}
  \Phi = \begin{pmatrix}
    a_1 & &&&&\\
    & a_1 &&&&\\
    && a_2 &&&\\
    &&& a_4 &&\\
    &&&& \ddots &\\
    &&&&& a_n
    \end{pmatrix}
\end{equation}
Around any of these vacua $\Phi$ is massive so it can be integrated
out of the low energy theory, which is then given by $\cN=1$ SYM with
the Higgsed gauge group $\prod_{i=1}^n U(N_i)$. The $N_i$ here count
how many eigenvalues of $\Phi$ are equal to $a_i$. It is conventional
and convenient to denote this Higgsing process
$U(N)\rightarrow\prod_{i=1}^n U(N_i)$.

This theory confines, and it is believed that the low energy degrees
of freedom of this theory after confinement, similarly to the case of
pure SYM, are given by the glueball superfields $S_i$, one for each of
the gauge factors:
\begin{equation}
  S_i = - \frac{1}{32\pi^2} \Tr_{SU(N_i)} W_{\alpha} W^{\alpha}
\end{equation}
with $W_{\alpha}$ the spinorial chiral superfield appearing in the
gauge kinetic action, having lowest component the gluino
$\lambda_{\alpha}$. As explained by Veneziano and Yankielowicz
\cite{Veneziano:1982ah}, for pure SYM the low energy dynamics can be
obtained from an effective superpotential involving the glueball
superfield:
\begin{equation}
  \label{eq:W-VY}
  W_{eff}(S) = S\log\left(\Lambda^{3N}/S^N\right) + NS
\end{equation}
where $\Lambda$ denotes the dynamical scale of the $SU(N)$ theory.

In the case of the theory we are dealing with, the effective
superpotential will be of the form \cite{Dijkgraaf:2002fc}:
\begin{equation}
  \label{eq:W-eff}
  W_{eff}(S) = \sum_{i=1}^n \left[S_i\log\left(\frac{\Lambda_i^{3N_i}}{S_i^{N_i}}\right)
    + N_iS_i + N_i \frac{\partial \cF_0(S)}{\partial S_i}\right],
\end{equation}
where $\cF_0$ is a perturbative series on the glueballs $S_i$, and
$\Lambda_i$ is the dynamical scale of the factor $SU(N_i)$ after
integrating out the massive matter. The usual matching relations lets
us express $\Lambda_i$ in terms of the dynamical scale $\Lambda$ of
the original theory, the derivative of the superpotential $W'(z)$, and
the residual ranks $N_i$:
\begin{equation}
  \Lambda_i^{3N_i} = \Lambda^{2N} g^{N_i} \prod_{j\neq i} (a_j - a_i)^{N_i-2N_j}
\end{equation}
Similar matching relations for the related $SO(N)\rightarrow
SO(N_0)\times \prod_i U(N_i)$ and $Sp(N)\rightarrow Sp(N_0)\times
\prod_i U(N_i)$ cases can be found in \cite{Intriligator:2003xs}.

What Dijkgraaf and Vafa showed is that $\cF_0$ can be computed
perturbatively from a related matrix model. We describe the main
points of their argument below, but let us first describe how we will
embed these gauge theories into string theory.

\subsection*{Geometry}

A type IIB configuration realizing the gauge theory we are studying
was described in \cite{Kachru:2000ih}. It is given by Minkowski
spacetime times a local Calabi-Yau defined by the resolution of:
\begin{equation}
  \label{eq:DV-geometry-W}
  x^2 + y^2 + w^2 + (W'(z))^2=0
\end{equation}
with $W(z)$ the same function as in (\ref{eq:W-adjoint}), and $W'(z)$
its first derivative with respect to $z$. The extrema of $W(z)$ are
located at $z=a_i$, so we can rewrite (\ref{eq:DV-geometry-W}) more
clearly as:
\begin{equation}
  \label{eq:DV-geometry}
  x^2 + y^2 + w^2 + g^2 \prod_{i=1}^n(z - a_i)^2=0.
\end{equation}

It is easy to see from here that when $z$ is close to $a_i$ (we take
the $a_i$ to be well separated for simplicity), the equation
(\ref{eq:DV-geometry}) reduces to the one for the conifold. We take
the conifold to be resolved\footnote{Resolution means, roughly, that
  we substitute the singularity of the conifold by a finite size ${\bf
    S}^2$ in such a way that the resulting space is smooth. See for
  example \cite{Candelas:1989js} for a more detailed description.}, so
the geometry is no longer singular, but instead has an isolated
2-cycle at each of the degenerations $a_i$. The size of the resolved
2-cycles is determined by the complexified gauge coupling constant for
the theory we are engineering, more on this below.

Let us consider this background in the presence of $N$ D5 branes,
which we take to wrap Minkowski and two internal directions. The low
energy dynamics of such a system is $U(N)$ with the superpotential
(\ref{eq:W-adjoint}) for the adjoints. Classical vacua of the system
correspond to distributing the $N$ D-branes in different ways in the
$n$ different 2-spheres of the resolved conifolds. When the $N$ branes
get separated into the different cycles, some open strings get
massive, and the gauge group on the branes factorizes. Furthermore the
adjoint chiral multiplet living on the brane, and which parametrizes
motions of the brane, gets a positive mass since the resolution
two-sphere is a local volume minimum in its homology class.

Let us briefly mention that all the resolved two spheres belong to the
same homology class. One simple way to see this in physics terms is
that we can tunnel from $U(N_1)\times U(N_2)$ to $U(N_1-1)\times
U(N_2+1)$ in the gauge theory just by smoothly taking one eigenvalue
of $\Phi$ from $a_1$ to $a_2$. For such a tunneling to be possible in
string theory the total homology class should stay the same, the
$a_1\rightarrow a_2$ interpolation describes the 3-chain that makes
the two cycles homologous. Such a tunneling is described by instantons
of the matrix model \cite{Dijkgraaf:2002fc,Marino:2007te}, which are
related to domain walls in the effective theory interpolating between
the two vacua, and coming from branes wrapping the 3-chain.

The bare holomorphic coupling constant $\tau$ of the Yang-Mills theory
living on the brane is given by the complexified volume of the two
cycle $S$ wrapped by the brane:
\begin{equation}
  \label{eq:tau}
  \tau = \frac{\theta}{2\pi} - \frac{4\pi i}{g_{YM}^2} =
  \frac{1}{2\pi i g_s} \int_S \left(J + iB\right).
\end{equation}

We will also be interested in engineering $Sp(N)$ theories with a
superpotential for the adjoint. They can be obtained by adding a
suitable orientifold to the geometry above. The breaking pattern in
this case is $Sp(N)\rightarrow Sp(N_1)\times U(N_2)\times \ldots
\times U(N_n)$, but otherwise everything is very similar to the $U(N)$
case \cite{Intriligator:2003xs}.

\subsection*{The low energy degrees of freedom of the string theory}

We see that up to now the tree level dynamics of the stringy system
and the gauge theory are very similar, and can be matched quite
easily. Nevertheless, we have not yet described what plays the role of
the all-important glueball superfield in the string side. This is a
subtle problem, and central to our discussion.

It turns out that in order to account for all the low energy dynamics
of string theory one needs to include more ``glueball'' superfields in
the string theory side than those that would be necessary just from
the gauge theory point of view. In fact, one way of reading the
results of this note is that these extra degrees of freedom in the
string side are those necessary in order to account for the stringy
instanton effects, which are not present in gauge theory.

These extra glueballs must be included for the cases in which there is
a single brane wrapping a ${\bf S}^2$, giving rise to a $U(1)$ gauge
group, and in the case on which the ${\bf S}^2$ is wrapped by no
branes, but there is an orientifold such that the gauge group on the
cycle, had there been $N$ branes wrapped on it, would have been
$Sp(N)$. We will refer to this last case as the $Sp(0)$ case. In gauge
theory, none of this two gauge groups would give rise to a glueball
and a Veneziano-Yankielowicz superpotential describing its
dynamics. These are also the only cases on which one has to include
these extra degrees of freedom in order to match with the
superpotential due to one stringy instanton effects. Other candidates
one could think of such as $SO(0)$, $SO(2)$ or $U(0)$ do not require
extra glueballs for describing the stringy dynamics. The justification
of this result from our point of view is that these are the D-brane
instantons that contribute to the superpotential (see
Section~\ref{sec:instanton-review} below), so we need to make the
corresponding glueballs dynamic in order to reproduce the known
stringy instanton effects.

Note that the prescription given in \cite{Intriligator:2003xs} for
when to include the extra stringy glueballs, which was obtained via
the geometric transition, coincides with our prescription that the
matrix model should compute D-brane instanton effects, which is a
successful check of our proposal.

We should point out that the previous discussion is the correct one
when we are restricting ourselves to superpotential contributions in
the single instanton case (as we are doing in this note), but we
expect that it might be necessary to consider dynamical at least the
$U(0)$ glueballs in the description of more complicated processes. The
reason is the following. Consider instantons in a $U(0)$ gauge group
(these are commonly called $U(1)$ instantons in the D-brane instanton
literature). They have four neutral zero modes, and thus they do not
contribute to the superpotential by themselves, in agreement with the
fact that we are considering the corresponding glueball
non-dynamical. Nevertheless, as discussed in
\cite{GarciaEtxebarria:2007zv}, multi-instanton processes involving a
$Sp(0)$ and a $U(0)$ instanton can contribute to the
superpotential. We cannot reproduce these effects in the matrix model
side if we just set the $U(0)$ glueball to zero from the
beginning. One way out is to include the $U(0)$ glueball in the game,
and just set to zero the Veneziano-Yankielowicz part of the
superpotential. Namely, we take the following formula to hold for any
$N\geq 0$:
\begin{equation}
  \label{eq:U(N)-bare}
  W_{U(N)} = S\log\left(\Lambda_0^{3N}/S^N\right) + NS - 2\pi i\tau S
\end{equation}
Here $\Lambda_0$ is the cutoff for the theory, which combines with the
bare holomorphic coupling $\tau = \frac{\theta}{2\pi} - \frac{4\pi
  i}{g_{YM}^2}$ in order to give the usual holomorphic scale $\Lambda$
of the gauge theory:
\begin{equation}
  \Lambda^3 = \Lambda_0^3 \, e^{-2\pi i \tau/N}
\end{equation}
reproducing in this way the Veneziano-Yankielowicz superpotential
(\ref{eq:W-VY}).

Note that the first two terms in (\ref{eq:U(N)-bare}) vanish when
$N=0$. For the one-instanton superpotential, such a prescription
dynamically sets $S=0$ due to the effect of an extra light
hypermultiplet appearing close to $S=0$ \cite{Intriligator:2003xs},
but for multi-instanton processes there could be nontrivial
contributions.

This prescription for the $U(0)$ glueball has a chance of matching the
multi-instanton results, and also fits well with the heuristic idea
that dynamical glueballs are associated with instanton effects in the
corresponding node. We hope to come back to this interesting question
in the future. In any case, we will not go beyond the one-instanton
sector in this note, so this subtlety will not be important and we
refrain from further discussion of this point.

\subsection*{Matrix models as the theory on the branes}

Let us now come back to the connection between the string theory and
the matrix model, following
\cite{Dijkgraaf:2002fc,Dijkgraaf:2002vw,Dijkgraaf:2002dh}. Since we
are interested in computing F-terms, we can go to the topological
string. As described above, we have $N$ D5 branes in a geometry
defined by the function $W(z)$. The topological theory living on the
worldvolume of the D5 branes is the reduction of three complex
dimensional holomorphic Chern-Simons theory \cite{Witten:1992fb} to
the two dimensional worldvolume of the brane. Partially solving the
equations of motion for this theory tells us that the F-term physics
is captured by the partition function of the holomorphic matrix model
with action:
\begin{equation}
  \label{eq:Z-mm}
  \cZ = \frac{1}{\mathrm{Vol}(U(M_1)\times \cdots \times U(M_n))} \exp
  \left(- \frac{1}{g_s} \int d\Phi \, \Tr W(\Phi)\right)
\end{equation}
This formula requires some explanation. Here $\Phi$ is a $M\times M$
matrix ($M$ is not related to the number of branes $N$, its relevance
to the physical theory will be explained shortly), and $g_s$ is a
overall scale unrelated to the coupling constant of the physical
theory. As above, the classical vacua of this theory will be
determined by extremizing $W(\Phi)$, and distributing the $M$ vacua
into the $n$ extrema. We take $M=M_1+M_2+\ldots+M_n$. The prefactor
takes into account the volume of the unbroken gauge group in the
chosen vacuum.

For the particular case of the superpotential, which will be the main
focus of this note, it can be further argued \cite{Bershadsky:1993cx}
that only the planar diagrams contribute\footnote{The argument boils
  down to the fact that the planar diagrams are the ones without
  momentum insertions. It also implies that the higher F-terms, which
  do involve momentum insertions, are computed by the non-planar part
  of the matrix model. We will come back to this observation in the
  conclusions.}. This means that the required information can be
obtained in the large $M$ limit of the matrix model.  As usual when
taking large $M$ limits, we will keep the 't Hooft coupling
$S_i=g_sM_i$ fixed. This has two desirable consequences. First it
tells us that we can solve the matrix model in the saddle point
approximation, which is easier; and second, one can apply the
arguments of \cite{Vafa:2000wi,Cachazo:2001jy} to argue that the
result can be computed via a geometric transition to a flux
background.

\medskip

Let us now describe how the prescription goes for connecting the
matrix model amplitudes to F-terms in the physical string. The 't
Hooft parameters $S_i$ of the matrix model get identified with the
glueballs $S_i$ of the physical string. In this way, the perturbative
expansion of the matrix model is an expansion in powers of the
glueballs. The conjecture is that the low energy superpotential of the
string theory is given by:
\begin{equation}
  \label{eq:W-eff-mm}
  W_{eff}(S) = \sum_{i=1}^n\left[N_i \frac{\partial \cZ_0(S)}{\partial
      S_i} - 2\pi i \tau_i S_i \right]
\end{equation}
where $N_i$ denotes the rank of the gauge factor $U(N_i)$ and $\tau_i$
is the bare holomorphic gauge coupling on the branes wrapping the
cycle which is determined by the volume of the wrapped two cycle as in
eq. (\ref{eq:tau})\footnote{When computing instanton effects,
  sometimes there will be no spacetime filling D-brane wrapping the
  cycle, and thus the definition of $\tau$ as a gauge coupling
  constant does not make sense. The definition in terms of the
  (complexified) volume of the resolved cycle still makes sense, and
  it is the one we will always be implicitly using.}. $\cZ_0$ denotes
the planar part of the matrix model partition function
(\ref{eq:Z-mm}).

Expressions (\ref{eq:W-eff-mm}) and (\ref{eq:W-eff}) match nicely once
we take into account the following. As explained in
\cite{Ooguri:2002gx} from the mirror Chern-Simons perspective, the
answer for the free planar part of the matrix model, once we take the
volume of the residual gauge group into account, gives the Veneziano
Yankielowicz superpotential. We can then expand the planar partition
function into a free part and a correction:
\begin{equation}
  \label{eq:cZ_0-expansion}
  \cZ_0 = \left\{\frac{1}{\mathrm{Vol}(U(M_1)\times \cdots \times U(M_n))}
  \exp \left(- \frac{1}{g_s} \int
    d\Phi \, \Tr \frac{\Phi^2}{2} \right)\right\} + \cF_0
\end{equation}
The first term gives rise to the Veneziano-Yankielowicz term in the
superpotential (notice also that it is the partition function for the
conifold), and $\cF_0$ is a polynomial in $S_i$:
\begin{equation}
  \cF_0 = \sum_{i_1\ldots i_n} c_{i_1\ldots i_n} S_1^{i_1}\cdots S_n^{i_n}
\end{equation}
This series expansion in $S_i$ can be understood as a systematic
fractional instanton expansion, where the power of $S_i$ determines
the number of fractional instantons in the factor $U(N_i)$ we are
taking into account. In this note we will be considering just the one
instanton sector, which is determined by the universal term coming
from the free (conifold) theory.  We leave the study of
multi-instanton effects due to the $\cF_0$ expansion for future work.

\subsection{D-brane instantons}
\label{sec:instanton-review}

Let us now proceed to review the aspects of D-brane instanton calculus
most relevant to our discussion. As reviewed in the previous section,
the Calabi-Yau has cycles over which we can wrap space-filling
D-branes in order to engineer 4d supersymmetric gauge
theories. Depending on the particular vacuum we choose there might
also be cycles not wrapped by any spacetime filling D-brane. Over
these cycles we might wrap D-brane instantons, which in this context
are branes wrapping these internal cycles and localized at a point in
spacetime. They give nonperturbative (in $g_s$) contributions to the
4d effective action living on the spacetime wrapping branes.

The basic condition for a D-brane instanton to contribute to the
superpotential in this setting is that it has exactly two unlifted
fermionic zero modes, which will play the role of the $\theta$
variables in the integration of the superpotential over
superspace. The kind of backgrounds where are considering preserve, in
the absence of branes and orientifolds, 8 supercharges. A D-brane
instanton is a 1/2 BPS object, and thus has 4 fermionic zero modes
$\theta_\alpha$, $\bar \theta_{\dot \alpha}$ corresponding to the 4
spontaneously broken supersymmetries. If the D-brane instanton is
going to contribute to the superpotential, we should include some
mechanism that lifts the two $\bar \theta_{\dot \alpha}$ Goldstinos.

The simplest configurations in which this happens, and which have no
simple gauge theory interpretation, are instantons of the $U(1)$ and
$Sp(0)$ types. We describe them in the following.

\medskip

An instanton of $Sp(0)$ type occupies a cycle in which there are no
wrapped space-filling branes, and which is mapped to itself by the
orientifold action. Furthermore, the sign of the orientifold
projection is such that the gauge group on the instanton is
$O(1)$. Such a projection does indeed remove the $\bar \theta_{\dot
  \alpha}$ Goldstinos, as we want\footnote{See for example
  \cite{Ibanez:2007rs} for a nice discussion of the details of the
  possible orientifold actions and their effect on the instanton zero
  modes}.

In order to make the connection with the matrix model more
transparent, here we have chosen to refer to this kind of instanton as
``$Sp(0)$'', instead of using the more common denomination
``$O(1)$''. The justification for this denomination from the instanton
point of view is the following. Consider the orientifold action on $N$
space-filling D-branes wrapping the same cycle as the instanton. Since
the orientifold projection acts oppositely on space-filling and
instanton branes, these branes would have gauge group $Sp(N)$. The
D-brane instanton in this context can be interpreted as the gauge
instanton for $Sp(N)$, with the couplings for the string modes between
the different branes implementing the ADHM construction
\cite{Billo:2002hm,Akerblom:2006hx,Bianchi:2007fx,Argurio:2007vqa,Bianchi:2007wy}.
The notation $Sp(0)$ is thus intended to suggest that the D-brane
instanton can be thought of as a gauge instanton of a (admittedly
rather empty) $Sp(0)$ gauge group.

\medskip

Let us now describe instantons of the $U(1)$ type. These instantons do
not exist in pure field theory either, but they do exist in string
theory when there is a single D-brane wrapping the same cycle as the
instanton
\cite{Aganagic:2007py,GarciaEtxebarria:2007zv,Petersson:2007sc}.

In order to explain why this happens, let us consider first $N$
spacetime filling branes wrapping the cycle, giving a $U(N)$ gauge
theory. Intuitively, since it is on top of the D-branes, the D-brane
instanton ``feels'' just $\cN=1$, and thus it has the two desired
Goldstinos $\theta_{\alpha}$ only. More precisely, the couplings
between the zero modes of the instanton and the fields living in the
branes implement the ADHM constraints, and the ADHM couplings lift the
$\bar \theta_{\dot \alpha}$ modes.

One interesting observation one could make is that the way in which
the couplings between the zero modes implement the ADHM constraints
and lift the two extra goldstinos does not require that $N$ is bigger
than 1 for $U(N)$, it works for $U(1)$ gauge theories too as studied
in detail in \cite{Petersson:2007sc}. $U(1)$ theories do not have
instanton effects in gauge theory, and the fact that D-brane
instantons contribute in this case is very much related to the fact
that we are completing this $U(1)$ theory in the UV by embedding into
string theory. To restate one of the main points of
\cite{Intriligator:2003xs} and also of this work: such a completion is
not as innocuous as it looks, since it also changes the low energy
superpotential due to these exotic $U(1)$ instantons.

\medskip

Before going on to check the agreement between the matrix model and
the D-brane instanton calculus in these cases, let us make a couple of
important remarks. Although we restrict here to the case in which the
instanton has exactly two zero modes, the general case in which the
instanton has more zero modes is also interesting. What happens here
is that the instanton does not contribute to the superpotential, but
can generate couplings involving more than two fermions, called higher
F-terms \cite{Beasley:2004ys,Beasley:2005iu}. One important example is
the gauge instanton for SQCD with $N_f=N_c$. These instantons do not
contribute to the superpotentials, but induce operators with
insertions of more than two fermions (or equivalently,
derivatives). They are of the form:
\begin{equation}
  \int d^4x d^2\theta \, w_{\overline i_1\overline j_1\ldots
    \overline i_n\overline j_n} (\Phi, \overline \Phi) \,\,
  \left({\overline D} {\overline \Phi}^{\overline i_1} \cdot
    {\overline D} {\overline \Phi}^{\overline j_1} \right) \cdots
  \left({\overline D} {\overline \Phi}^{\overline i_n}\cdot {\overline D}
    {\overline \Phi}^{\overline j_n}\right)
\end{equation}
where $\Phi$ represents the moduli of the system, and $\omega$ is a
section of $\ov \Omega^p_\cM\otimes \ov \Omega^p_\cM$, with $\cM$ the
moduli space\footnote{In other words, it is a form antisymmetric in
  the $\overline i$ indices, antisymmetric in the $\overline j$
  indices, and these indices live in the cotangent bundle over the
  moduli space.}. We include some preliminary remarks on the relation
between matrix models and higher F-terms in the conclusions.

Another interesting possibility for lifting extra zero modes which we
have not discussed here is by using fluxes, as studied in
\cite{Blumenhagen:2007bn,GarciaEtxebarria:2008pi,Billo':2008sp,Billo':2008pg,Uranga:2008nh}.
Modifying the matrix model side of the story in order to incorporate
the effect of these fluxes is a worthwhile and interesting problem
which we will not attempt to solve here.

\section{The glueball superpotential for low rank and D-brane
  instantons}
\label{sec:low-rank}

In this section we will study the two basic cases in which the matrix
model result does not match the gauge theory result, namely, as
reviewed in Section~\ref{sec:DV-review}, configurations with gauge
group $U(1)$ and $Sp(0)$. We will see that there is a nice agreement
between the result of the D-brane instanton calculation and the result
obtained from the matrix model prescription. This gives evidence for
the advertised result that matrix models can be used to compute
D-brane instanton effects.

\medskip

The geometry we will be focusing on will be the resolved conifold, so
let us take a quadratic superpotential for the gauge theory:
\begin{equation}
  \int d^2\theta \, \Tr W(\Phi)=\int d^2\theta \, \Tr \frac{\Phi^2}{2}
\end{equation}
where the factor of $1/2$ is for later convenience. Plugging this
$W(x)$ into equation (\ref{eq:DV-geometry-W}) we obtain the conifold
equation $x^2+y^2+z^2+w^2=0$. As explained above, we will be
considering the resolved conifold, with a resolution parameter
determined by the theory we are engineering.

\subsection*{The matrix model computation}

Let us start by the matrix model side of the story, which is the
simplest. We have to compute the planar limit of the matrix model free
energy in this background. The conifold is described by the free
holomorphic one-matrix model, so we just obtain the
Veneziano-Yankielowicz result. Note that we must set $N=1$ in
eq.~(\ref{eq:W-eff}) in order to capture $Sp(0)$ and $U(1)$ D-brane
instanton effects, as discussed in Section~\ref{sec:DV-review}. We
get:
\begin{equation}
  \label{eq:Sp(0)-Weff-S}
  W_{eff}(S,\tau) = S\log \left(\frac{\Lambda_0^3}{S}\right) + S -2\pi i \tau S
\end{equation}
with $\Lambda_0$ the cutoff scale and $\tau$ the bare coupling of the
configuration, defined in eq.~(\ref{eq:tau}):
\begin{equation}
  2\pi i \tau = \frac{1}{g_s}\int_{{\bf S}^2} J + iB
\end{equation}
We can now proceed to integrate $S$ out. Its equation of motion gives:
\begin{equation}
    S = e^{-2\pi i\tau}\Lambda_0^3
\end{equation}
which upon substitution into (\ref{eq:Sp(0)-Weff-S}) gives:
\begin{equation}
  \label{eq:MM-Sp(0)-Weff}
  W = e^{-2\pi i \tau}\Lambda_0^3 = \Lambda^3
\end{equation}
where we have introduced the cutoff independent scale $\Lambda$.

\subsection*{The D-brane instanton computation}

Let us now describe in some detail how the stringy instanton result
reproduces this contribution. In order to avoid the complications of
having to impose the (super)ADHM constraints, we will focus on the
$Sp(0)$ case, so the $\bar \theta$ modes are simply projected out by
the orientifold. A good place to see how the saturation of the $\bar
\theta$ modes goes for $U(1)$ instantons is for example
\cite{Petersson:2007sc}.

We will start by considering the case with $W=0$. In this case we will
have 8 background supercharges close to the orientifold, and 16 in the
bulk. The $Sp(0)$ instanton is a 1/2 BPS object wrapping a 2-cycle
mapped to itself under the orientifold action, so it will have 4
Goldstinos in its worldvolume. Let us call the surviving Goldstinos
$\theta^A_\alpha$, where $A$ is an index taking values in the
fundamental representation of the $SU(2)$ R-symmetry group preserved
by the orientifold. Such an instanton has four neutral zero modes, and
thus cannot contribute to the superpotential.

We now switch on the quadratic superpotential $W(z)=\frac{1}{2}m
z^2$. This reduces the supersymmetry of the background by half, so now
we have 8 supercharges on the bulk and 4 close to the orientifold. In
such a configuration it is reasonable to guess that two of the four
zero Goldstinos of the instanton get lifted, and the instanton can now
contribute to the superpotential. This is indeed the case, let us now
explain how this deformation acts from the point of view of the
effective action on the instanton zero modes.

Let us start by studying the case were we have background spacetime
filling branes, so we are dealing with ordinary gauge theory
instantons. As reviewed in Section~\ref{sec:DV-review}, in the gauge
theory the deformation of the background appears as a breaking of the
supersymmetry from $\cN=2$ to $\cN=1$ by the addition of the
superpotential for the adjoint chiral multiplet:
\begin{equation}
  W(\Phi) = \frac{1}{2} m \Tr \Phi 
\end{equation}
One consequence of such a coupling is that some of the zero modes
associated to the field $\Phi$ get lifted, let us call such modes
$\lambda_{\alpha}$ and $\bar \lambda_{\dot \alpha}$. We have chosen
this notation in order to remind the reader that the adjoint
superfield comes from the vector multiplet of $\cN=2$, so the highest
component fermion of the multiplet can be thought of as a gaugino from
the point of view of the broken $\cN=1$ algebra. As reviewed in
Section~VI.4 of \cite{Dorey:2002ik}, this can be nicely encoded in the
effective action for the instanton zero modes as the addition of the
following term:
\begin{equation}
  \label{eq:lambda-mass}
  \delta S_{inst} = -\frac{m\pi^2}{g_{YM}} \lambda^\alpha \lambda_{\alpha}
\end{equation}
Such a term can be used to saturate the $d^2\lambda$ term in the
instanton measure, allowing the instanton to contribute to the
superpotential.

We now claim that a term similar to (\ref{eq:lambda-mass}) also exists
for our $Sp(0)$ instanton. We cannot compute the existence of such a
coupling via CFT methods in string theory, but given the fact that
string theory gives rise to the whole ADHM construction in such a
natural way \cite{Witten:1995gx, Douglas:1996uz}, with a natural
identification of the instanton-instanton strings as gauge theory zero
modes, assuming that an analog of (\ref{eq:lambda-mass}) holds seems
to be reasonable.

From the point of view of the instanton brane $\lambda_{\alpha}$ is
one of the Goldstinos $\theta^A_{\alpha}$, let us take it to be
$\theta^2_{\alpha}$. The coupling induced by the deformation would
then be:
\begin{equation}
  \delta S_{inst} = -\frac{m\pi^2}{g_{YM}} \theta^{2\alpha} \theta^2_{\alpha}
\end{equation}
Just as in the gauge theory case, we can bring down this coupling in
order to saturate the fermionic integration over the $\theta^2_\alpha$
modes. This integration will give us a factor of $m$ in front of the
resulting contribution to the superpotential (the factors of $\pi$ and
$g_{YM}$ get taken care of once we integrate over the rest of the zero
modes properly \cite{Dorey:2002ik}).

The rest of the calculation, once we have dealt with the massive
adjoint, is by now well understood (see for example
\cite{Petersson:2007sc} for a careful derivation for the more involved
$U(1)$ instanton). The end result is simply given by:
\begin{equation}
  \label{eq:W-instanton}
  W_{eff}(\tau) = m M_s^2 \, \exp\left(-2\pi i \tau\right)
\end{equation}
The $M_s$ factors come from the measure of integration of the zero
modes \cite{Billo:2002hm,Bianchi:2007wy,Petersson:2007sc}, and the
exponential suppression comes from the volume of the cycle wrapped by
the instanton.

In order to match with the matrix model result we need to relate the
low energy physical $\Lambda$ appearing in (\ref{eq:MM-Sp(0)-Weff})
with the microscopic scales appearing in (\ref{eq:W-instanton}). This
goes as follows. The physical scale for the theory before integrating
out the adjoint is given in terms of the string scale quantities as:
\begin{equation}
  \Lambda^2_{adj} = M_S^2 e^{-2\pi i \tau}
\end{equation}
The $m$ prefactor gives the matching relation between the theory
before and after integrating out the adjoint:
\begin{equation}
  \Lambda^3 = \Lambda^2_{adj} m
\end{equation}
The resulting superpotential can thus be written as:
\begin{equation}
  W=\Lambda^{3}
\end{equation}
matching (\ref{eq:MM-Sp(0)-Weff}) beautifully.

\section{A quiver example: $\bZ_n$ orbifold of the conifold}
\label{sec:quiver}

Let us proceed to the more interesting case in which the D-brane
instanton intersects some space-time filling branes. In this case, as
discussed in \cite{Blumenhagen:2006xt,Ibanez:2006da,Florea:2006si},
there are fermionic zero modes arising from strings stretching between
the instanton and the brane. Integration over them typically induces
interesting couplings in the world volume theory of the brane, which
might be potentially relevant for model building. They generically
give rise to couplings with are forbidden in perturbation theory.

In this section we will focus on a model which is easy to analyze
using matrix model techniques. It is the theory corresponding to
branes at the singularity of an abelian orbifold of the
conifold. These theories can be easily obtained from $A_n$ orbifold
quivers by giving large masses to the adjoints.  The analysis we do
here is easily generalizable to $\cN=2$ quiver theories with
superpotentials for the adjoints, as in the original paper of
Dijkgraaf and Vafa \cite{Dijkgraaf:2002vw}. We will choose an
assignment of ranks of the quiver such that the orientifolded node is
empty, giving rise to one of the $Sp(0)$ factors we have been
describing.

\FIGURE[t]{
  \includegraphics[width=0.8\textwidth]{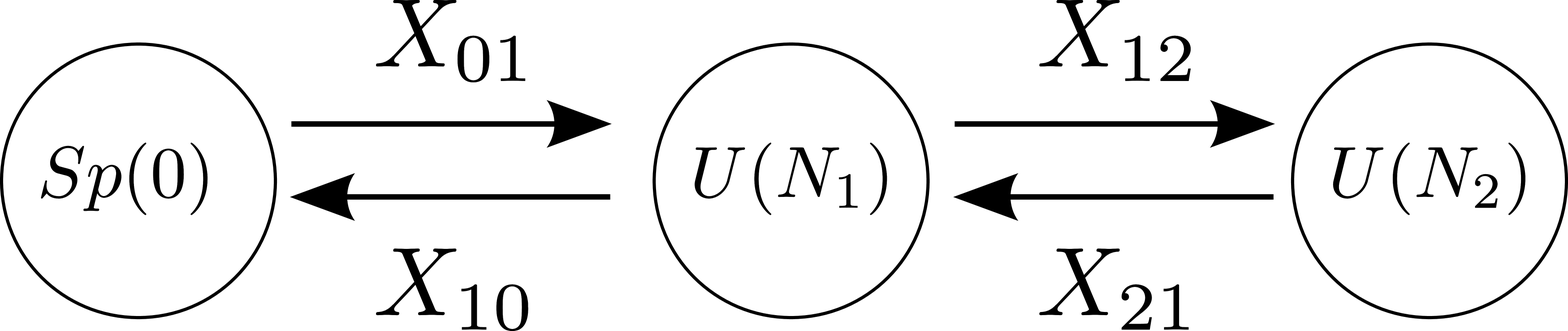}

  \caption{Quiver for the orientifolded orbifold of the conifold we
    discuss in the main text. We have omitted the part of the quiver
    to the right of $U(N_2)$, which will not be relevant for our
    discussion.}
  \label{fig:sp(0)-conifold}
}

The relevant quiver diagram is shown in
Figure~\ref{fig:sp(0)-conifold}, where we are omitting the part of the
quiver that will not be relevant for us. The nodes of the diagram
denote gauge factors, and the arrows denote bifundamental chiral
multiplets. The theory also has a quartic superpotential, given for
generic ranks in the nodes by:
\begin{equation}
  \label{eq:sp(0)-W}
  W = \Tr X_{01} X_{12} X_{21} X_{10}
\end{equation}

In our case, since we are setting the rank of the $Sp$ node to zero,
$X_{01}$ and $X_{10}$ vanish, and $W=0$. Nevertheless, as we saw in
Section~\ref{sec:DV-review}, the ranks that go into the matrix model
computation are unrelated to those in the physical string. This means
that we can compute the effective superpotential keeping $X_{01}$,
$X_{10}$ and $W$ in the game, and treating them as if we had $Sp(K)$
with $K>0$. The only place where the actual rank of $Sp(K)$ enters in
the matrix model computation is in determining the prefactor in the
Veneziano Yankielowicz term, as in eq. (\ref{eq:W-eff}).

\medskip

Let us proceed to compute the low energy superpotential from the
matrix model point of view. We want to obtain the effect induced by
one instanton in the $Sp(0)$ node. One way we could go about computing
this effect would be by applying the prescription in
\cite{Dijkgraaf:2002vw} for $A_n$ quivers and taking the limit where
the adjoint masses go to infinity, keeping the dynamical scales of the
resulting conifold theory finite. See for example
\cite{Seki:2002ti,Casero:2003gr,Chiantese:2003qb,Ookouchi:2002be} for
calculations along these lines. Since we are not interested in the
full answer, but just the leading term due to $Sp(K)$, we can take
instead the following shortcut.

The leading term we are interested in corresponds to letting $Sp(K)$
confine while taking the rest of the quiver theory to be an
spectator. In this way $U(N_1)\times U(N_2)$ can be taken to be a
flavor group, and $X_{12}$, $X_{21}$ are background vevs. From this
point of view, we can take (\ref{eq:sp(0)-W}) to mean:
\begin{equation}
  W = \Tr X_{01} X_{12} \langle X_{21} X_{12} \rangle
\end{equation}
In this equation $\langle X_{21}X_{12} \rangle$ is the mass matrix for
$X_{01}$, $X_{12}$, which we can now integrate out. It is important to
take into account the matching relation:
\begin{equation}
  \Lambda_{low}^{3(K+1)} = \Lambda_{high}^{3(K+1)-N_1} \det \langle X_{21} X_{12}
  \rangle
\end{equation}
where $\Lambda_{high}$ and $\Lambda_{low}$ are the dynamical scales of
the $Sp(K)$ factor before and after integrating out the massive
matter.

The remaining theory is $Sp(K)$ with no flavors. It confines, and the
low energy superpotential is given by:
\begin{equation}
  W = (K+1)\Lambda_{low}^{3} =
  (K+1)\Lambda_{high}^{3-N_1/(K+1)} \det \langle X_{21} X_{12}\rangle^{\frac{1}{K+1}}
\end{equation}

We can now use this result computed in the gauge theory with $K>0$,
extract the matrix model result for $\partial \cZ_0 / \partial S$, and
apply it to the $Sp(0)$ case. We obtain the result:
\begin{equation}
  \label{eq:W-coni-MM}
  W_{eff} = \Lambda^{3-N_1} \det \langle X_{21} X_{12}\rangle
\end{equation}

\medskip

Let us now compare this result with the D-brane instanton
computation. The technology for dealing with this problem is by now
standard \cite{Blumenhagen:2006xt,Ibanez:2006da,Florea:2006si}. An
instanton sitting on the $Sp(0)$ node has two neutral zero modes
$\theta_{\alpha}$, and two zero modes $\alpha$, $\beta$ (these are
anticommuting scalars living in the fundamental and antifundamental of
$U(N_1)$) arising from strings going from the instanton to the
$U(N_1)$ brane. Essentially the same physics \cite{Kachru:2008wt} that
gives rise to the superpotential (\ref{eq:sp(0)-W}) gives rise to the
following coupling in the instanton action:
\begin{equation}
  \delta S_{inst} = \alpha X_{12} X_{21} \beta
\end{equation}

The contribution of the instanton to the low energy effective
superpotential is then given by:
\begin{equation}
  W_{eff} = m M_s^{2-N_1} e^{-2\pi i \tau} \int d\alpha d\beta  \, \exp \left(
    \delta S_{inst}\right)
\end{equation}
The overall prefactor works exactly as in the conifold case studied in
Section~\ref{sec:low-rank}, and we can identify it with
$\Lambda^{3-N_1}$ above. The second term can be used to
saturate the integrations over $\alpha$ and $\beta$, and gives rise
to:
\begin{equation}
  \int d\alpha d\beta \, \exp \left(
    \delta S_{inst}\right) = \det X_{12} X_{21}
\end{equation}
reproducing the result (\ref{eq:W-coni-MM}) obtained from the matrix
model.

\medskip

We see that the essential ingredients that went into the previous
calculation are those that appeared already in
Section~\ref{sec:low-rank}, where we studied the case of the isolated
instanton. Namely, once we considered the glueball superfield for the
node containing the instanton dynamical, the matrix model correctly
gave us the same answer as the (in general more involved) instanton
calculation.

\section{Conclusions and further directions}
\label{sec:conclusions}

In this work we have started the study of D-brane instanton effects
from the point of view of matrix models. We found that at least the
simplest examples indicate that, in the same way that ordinary gauge
non-perturbative effects are captured by a perturbative calculation in
the matrix model, more stringy effects are also computed by the matrix
model. We checked this result in some detail for isolated conifolds,
and for quiver gauge theories admitting a matrix model description.

There are many ways in which one could extend the results in this
note, let us list here a few of them.

\medskip

We have restricted ourselves to matching the matrix model and the
D-brane instanton results just up to leading order in the instanton
expansion, namely the one (fractional) instanton sector. This
simplified the calculations enormously, reducing the problem to
writing down the Veneziano-Yankielowicz superpotential for the
instanton node and extremizing the complete superpotential. This is
already a good sign, and indicates that the matrix model formalism
might be significantly simpler to deal with (when it is applicable)
than the D-brane instanton calculus, which is sometimes involved even
in the one-instanton sector.

In this same line of argument, one technical difficulty to match the
higher orders in the matrix model perturbation expansion with
non-perturbative effects in string theory is that we will have to deal
with multi-(fractional) instanton effects in the string side, and
these are somewhat cumbersome to deal with using standard (D-brane)
instanton calculus. See however \cite{Dorey:2002ik} for a review of
some powerful techniques for dealing with multi-instantons in the case
of gauge theory with extended susy, \cite{Fucito:2005wc} for progress
in the $\cN=1$ case, and
\cite{Blumenhagen:2007bn,GarciaEtxebarria:2007zv,Blumenhagen:2008ji,Cvetic:2008ws,GarciaEtxebarria:2008pi,Camara:2008zk,Gaiotto:2008cd}
for some recent progress in more stringy systems. It would be
interesting to check the proposed correspondence for higher orders in
the instanton expansion using some of these techniques.

\medskip

The geometries on which one can apply directly the ideas of Dijkgraaf
and Vafa are rather restricted, and not very suitable for model
building\footnote{When we are interested just in the superpotential,
  we can study D-brane instanton contributions also via the geometric
  transition, as discussed in the introduction and
  Section~\ref{sec:DV-review}. There have been some interesting model
  building applications of these ideas recently, starting with
  \cite{Aganagic:2007py}.}. Since the time when the DV results
appeared there have been very interesting developments in the study of
matrix models and topological string theory. Particularly relevant for
our story is \cite{Bouchard:2007ys}, where a matrix-model inspired
formalism is given for computing perturbative amplitudes of the
B-model on a very rich set of geometric backgrounds (in particular,
this includes mirrors of toric Calabi-Yau spaces). It would be
extremely interesting to see if and how D-brane instanton effects
appear in this setup.

\medskip

We would also like to make some preliminary comments on the study of
higher F-terms in the context of matrix models (we have succinctly
introduced these higher F-terms towards the end of
Section~\ref{sec:instanton-review}). Recently it has become clear that
these higher F-terms form an integral part of the global picture for
D-brane instanton effects across moduli space
\cite{GarciaEtxebarria:2007zv,GarciaEtxebarria:2008pi,Uranga:2008nh}. They
have a fascinating interplay with ordinary superpotential terms in
order to give a consistent description of non-perturbative effects
across the whole moduli space of the compactification. It would be
rather interesting to see if and how these higher F-terms are encoded
in the matrix model description.

One seemingly reasonable possibility is that these terms are computed
by the non-planar diagrams in the matrix model. A naive and
straightforward reason is that we need to compute terms in the
effective superpotential with derivatives, or more than two fermions,
and it is well known that in the topological string higher genus
diagrams are associated with operators of this kind in the physical
superstring. This is obviously too sketchy, so it would be good to
have a more direct connection between non-planar diagrams and
Beasley-Witten higher F-terms. We know that non-planar diagrams encode
the effect of the C-deformation of the gluino anticommutation algebra
\cite{Ooguri:2003qp}, so if non-planar diagrams are also related to
higher F-terms then we would have a nice relation between
Beasley-Witten higher F-terms and the C-deformation.

\medskip

As a last remark, all of our discussion has been in the context of
type IIB string theory. By mirror symmetry, one could expect that
Chern-Simons theory in the appropriate background might also compute
the effect of exotic E2 instantons. It would be rather interesting to
see how this works.

\acknowledgments

I am happy to acknowledge interesting and fruitful discussions with
Marta Gomez-Reino, Daniel Krefl, Sergio Monta\~nez and Angel Uranga. I
also want to thank Nao Hasegawa for kind support and constant
encouragement. This work is supported by the High Energy Physics
Research Grant DE-FG05-95ER40893-A020.

\bibliographystyle{JHEP}
\bibliography{refs}

\end{document}